\documentclass[12pt,cite]{iopart}
\usepackage{iopams}  
\usepackage{graphicx,epsf}

\begin{document}

\title[Walking on fractals: diffusion and self-avoiding walks on percolation clusters]
{Walking on fractals: diffusion and self-avoiding walks on percolation clusters}

\author{V  Blavatska$^{1,2}$ and W Janke$^1$}

\address{$^1$ Institut f\"ur Theoretische Physik and Centre for Theoretical Sciences (NTZ),Universit\"at Leipzig, Postfach 100\,920,
D-04009 Leipzig, Germany}
\address{$^2$ Institute for Condensed Matter Physics, National
Academy of Sciences of Ukraine, UA--79011 Lviv, Ukraine}
\ead{viktoria@icmp.lviv.ua, Viktoria.Blavatska@itp.uni-leipzig.de}
\ead{Wolfhard.Janke@itp.uni-leipzig.de}

\begin{abstract}
We consider random walks (RWs) and self-avoiding walks (SAWs) on disordered lattices directly at the percolation threshold. 
Applying numerical simulations, we study the scaling behavior of the models on the incipient percolation cluster 
in space dimensions $d=2, 3, 4$. 
Our analysis yields estimates of universal exponents, governing the scaling laws for
configurational properties of RWs and SAWs.   
\end{abstract}

\pacs{67.80.dj, 36.20.-r, 64.60.ah, 07.05.Tp}
 
	
\maketitle
\section{Introduction}

The model of a random walk (RW)  provides a good description of
diffusion processes, such as for example  encountered for electrons in metals or colloidal particles in solution \cite{RWbook}.  The averaged mean square displacement of the 
diffusive particle at time $t$ (or, equivalently,
after $t$ steps on a lattice) scales as
\begin{equation} 
\langle R^2 \rangle \sim t^{2\nu_{{\rm RW}}}, \label{diff}
\end{equation}
where in a non-disordered medium $\nu_{{\rm RW}}=1/2$, independently of the space dimension $d$. A RW is a fractal object, with fractal dimension  $d_{{\rm RW}}=1/\nu_{RW}$.
The number of all possible trajectories $C_t$ for a randomly walking particle of $t$ steps can be found exactly: $C_t=z_0^t$, where $z_0$ is the coordination number 
of the corresponding lattice.

\begin{table}[b!]
\small{
\caption{ Critical concentration $p_c$ of site-diluted lattices and fractal
dimensions of percolation clusters, $d_{p_c}^F$, and the backbone of percolation clusters, $d_{p_c}^B$, 
for different space dimensions $d$.}
\label{dim}
\begin{indented}
\item[]
\begin{tabular}{@{}llll}
\br
 $d$ & $p_c$ & $d_{p_c}^F$ & $d_{p_c}^B$ \\ 
\mr
2 & $0.592746$ {\cite{Ziff94}}  & $91/49$ {\cite{Havlin87}}  & $1.650\pm0.005$ {\cite{Moukarzel98}}   \\ 
3 & $0.31160$ {\cite{Grassberger92}}  &  $2.51\pm0.02 $ {\cite{Grassberger86}} &  $1.86\pm0.01$ {\cite{Moukarzel98}}  \\ 
4 & $ 0.19688$ {\cite{Paul01}} & $3.05\pm0.05$ {\cite{Grassberger86}}  &  $1.95\pm 0.05 $ {\cite{Moukarzel98}}  \\ 
 \br
\end{tabular}
\end{indented}
}
\end{table} 

Forbidding the trajectory of a random walk to cross itself, we obtain a self-avoiding walk (SAW), which is one of the most successful 
in describing the universal configurational  properties of a long, flexible single polymer chain in good solvent \cite{desCloizeaux90}. 
The average squared end-to-end distance $\langle R^2\rangle$ and the number of configurations $ C_N  $  of SAWs with $N$ steps 
on the underlying lattice obey the scaling laws:
\begin{equation}\label{scaling}
 \langle R^2 \rangle
\sim N^{2\nu_{{\rm SAW}}},\mbox{\hspace{3em}}
 C_N  \sim z^{N}
N^{\gamma_{{\rm SAW}}-1},
\end{equation} 
where $\nu_{{\rm SAW}}, \gamma_{{\rm SAW}}$ are universal exponents that only depend on the
space dimensionality $d$, and $z$ is a non-universal fugacity, counting the average number of accessible nearest-neighbor sites.
  The properties of  SAWs on a regular lattice have been studied
in detail both in analytical approaches
\cite{Guillou80,Nienhuis82,Guillou85,Guida98} and computer simulations \cite{Rosenbluth55,Madras88,MacDonald92,MacDonald00,Li95,Caracciolo98}.  For example, in the
space dimension $d{=}3$ one finds within the frame of the field-theoretical renormalization group
approach  $\nu_{\rm SAW}{=}0.5882\pm 0.0011$ \cite{Guida98} and Monte
Carlo simulations give $\nu_{\rm SAW}{=}0.5877\pm0.0006$ \cite{Li95}.  For space dimensions $d$ above the upper
critical dimension $d_{\rm up}{=}4$, the effect of self-avoidance becomes irrelevant and SAWs behave effectively as random walks with exponents
$\nu_{{\rm RW}}=1/2$, $\gamma_{{\rm RW}}=1$.

The problem of random walks in disordered media is of great interest since it is connected with a large amount of physical phenomena: 
transport properties in fractures and porous rocks, the anomalous density of states in randomly diluted magnetic systems, silica aerogels 
and in glassy ionic systems, diffusion-controlled fusion of excitations in porous membrane films etc. (see, e.g., Ref. \cite{Havlin87} for a review).
 Similarly, SAWs on randomly diluted lattices  may serve as a model of linear polymers in a porous medium.

Much of our understanding of disordered systems comes from percolation theory \cite{Stauffer}. A disordered medium can be modelled as randomly diluted lattice, 
with a given concentration $p$ of lattice sites allowed for walking. Most interesting is the case, when 
$p$ equals the critical concentration $p_{c}$, the site-percolation threshold (see Table~\ref{dim}), 
and an incipient percolation cluster can be found in the system. 
Studying properties of percolative lattices, one encounters two possible statistical averages. 
In the  first, one considers  only percolation clusters with linear size much larger than the typical length of the physical phenomenon under discussion.  
The other statistical ensemble includes all the clusters which can be found in a percolative lattice. 
For the latter  ensemble of all clusters, the walks can start on any of the clusters, and for an $N$-step walk, performed on the $i$th cluster,
 we have $\langle R^2 \rangle \sim l_i^2$, 
where $l_i$ is the averaged  size of the $i$th cluster. 
In what follows, we will be interested in the former case, when trajectories of walks reside only on the percolation cluster.  
In this regime,  the scaling laws (\ref{diff}), (\ref{scaling})
hold with  new exponents $\nu_{{\rm RW}}^{p_c}\neq \nu_{{\rm RW}}$ \cite{Sahimi83,Majid84,Pandey83,Alexander82,Avraham82,Havlin83,Argyrakis84,McCarthy88,Lee00,Bug86,Hong84,Mastorakos93,
Webman81,Gefen83,Rammal83,Mukherjee95},
 $\nu_{{\rm SAW}}^{p_c}\neq\nu_{{\rm SAW}},\gamma_{{\rm SAW}}^{p_c}\neq\gamma_{{\rm SAW}} $ 
\cite{Kremer81,Lee89,Kim90,Woo91,Grassberger93,Lee96,Meir89,Lam90,
Nakanishi92,Rintoul94,Ordemann00,Nakanishi91,Barat91,Sahimi84,Rammal84,Kim87,
Roy90,Roy87,Aharony89,Lam84,Blavatska04,Janssen07,Blavatska08}.
 A hint
to the physical understanding of this phenomenon is given by the
fact that weak disorder does not change the dimension of a lattice, whereas the percolation cluster itself
is a fractal object with fractal dimension $d_{p_c}^F$ dependent on $d$ (see Table~\ref{dim}). 
In this way, scaling law exponents of residing walks change  with the
dimension $d_{p_c}^F$ of the (fractal) lattice on which the walk
resides. Since $d_{\rm up}{=}6$ for percolation
\cite{Stauffer}, the exponents $\nu_{{\rm SAW}}^{p_c}(d\geq 6){=}1/2$, $\gamma_{{\rm SAW}}^{p_c}(d\geq 6){=}1$.  

Our present paper aims to supplement the studies of random and self-avoiding walks on percolative lattices by obtaining numerical 
values for exponents, governing the scaling behavior of the models, up to $d=4$ by computer simulations. The layout of the paper is as follows: in the next section, we will present in detail the procedure of extracting the percolation cluster and its backbone on disordered 
lattices at the percolation threshold.  In section III we describe the pruned-enriched Rosenbluth algorithm, applied to study the scaling of self-avoiding walks, and present the results obtained. In the next section we  consider the method for studying random walks on percolation clusters. In Section V, we end up by giving conclusions and an outlook.

\section{Construction of percolation cluster}

\begin{figure}[b!]
\begin{center}
\includegraphics[width=4cm]{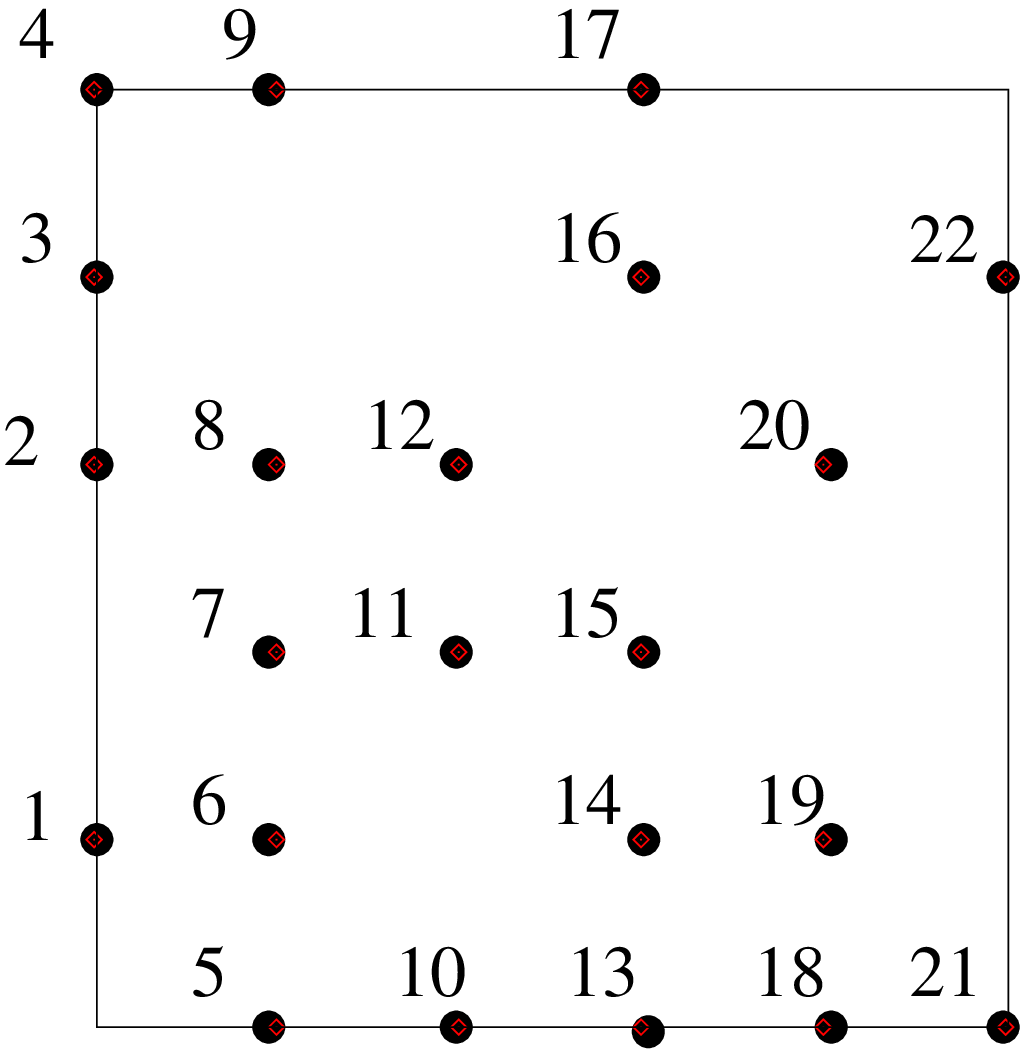}
\hspace*{2cm}
\includegraphics[width=3.8cm]{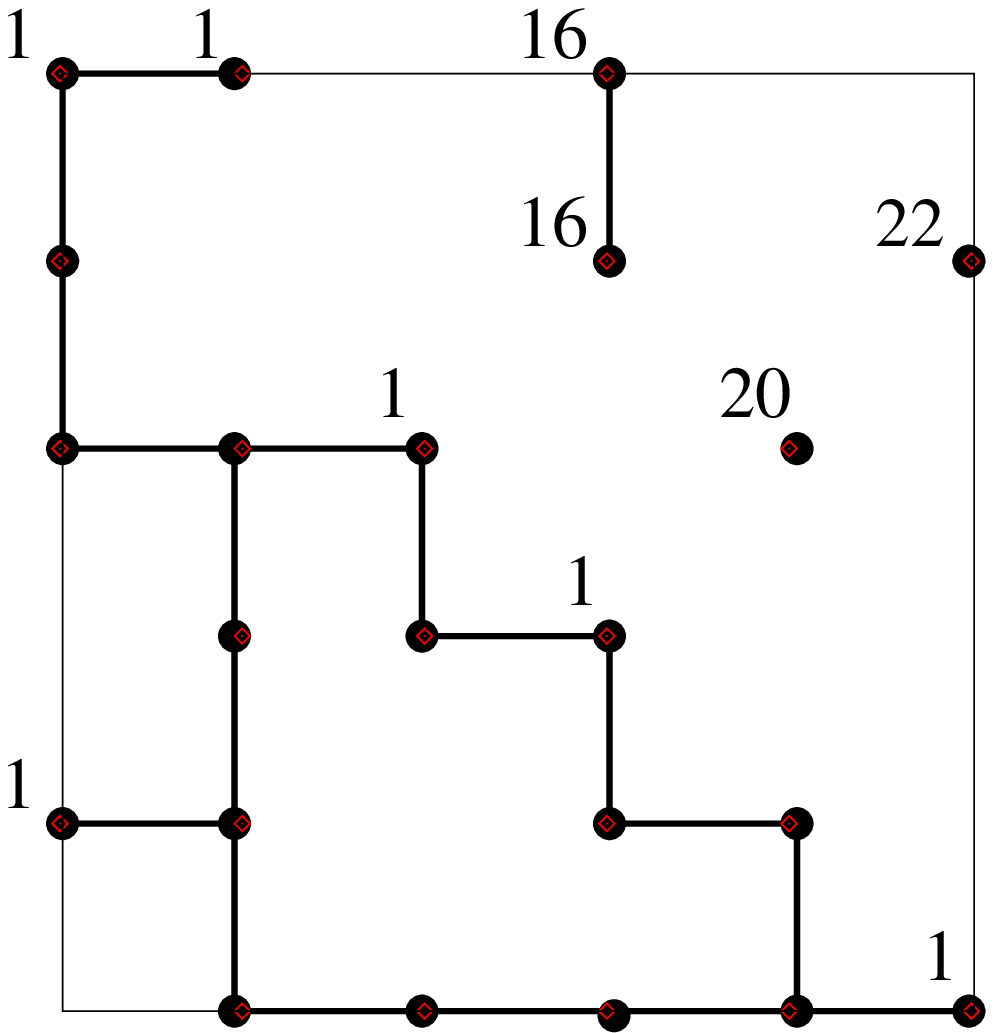}
\end{center}
\caption{\label{gratka}Procedure of site labeling and extracting the percolation cluster.}
\end{figure}

We consider site percolation on a regular lattice of edge length $L=400,200,50$ in dimensions $d=2,3,4$, respectively. Each site of the lattice is assigned to be 
occupied with probability $p_c$ (values of critical concentration in different dimensions are given in the Table~\ref{dim}), and empty otherwise. 
To describe the procedure of extracting the percolation cluster, let us consider schematically the two-dimensional case. We apply an algorithm based on the one proposed 
by Hoshen and Kopelman \cite{Hoshen76}. As a first step, a label is prescribed to each of the occupied sites.
Such a labeling process is regulated, we start, for example, from the first ``column" of the lattice, label the occupied sites upwards, and then turn 
to the next ``column", as shown in Fig.~\ref{gratka}, left.  Next, we start the procedure of burning the occupied sites. Again, in the same order, 
starting from the bottom of the first ``column" of the lattice, for each of the labeled sites (say, $i$), we check whether its nearest neighbors are 
also occupied or not. If yes, two possibilities appear:
1) The  label of the neighbor is larger than the label of site $i$. In this case, we change the label of the  neighbor to that of site $i$.
2) The label of the neighbor is smaller than that of $i$. Then, we change the label of site $i$ to that of the neighbor.

\begin{figure}[t!]
\begin{center}
\includegraphics[width=4cm]{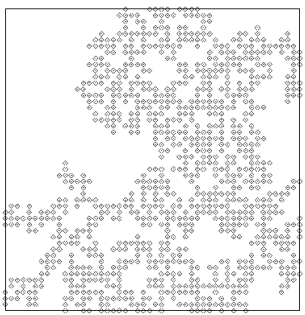}
\end{center}
\caption{\label{perc}Percolation cluster on a $d=2$-dimensional regular lattice of edge length $L=50$.}
\end{figure}

Such a burning procedure is applied until no more change of site labels is needed. As a result, we end up with groups of clusters of occupied 
sites with the same labels (Fig.~\ref{gratka}, right). Finally, we check, whether there exists a cluster, that wraps around the lattice in all $d$ directions. If yes, we have found the percolation cluster 
(Fig.~\ref{perc}). If not, this disordered lattice is rejected and a new one is constructed. Note, that on finite lattices the definition of spanning clusters is not unique (e.g.,
one could consider clusters connecting only two opposite borders), but
all definitions are characterized by the same fractal dimension and are
thus equally legitimate. The here employed definition has the advantage of
yielding the most isotropic clusters.
Note also that directly at $p=p_c$ more than one spanning cluster can be found in the system, and the probability $P(k)$ for at least $k$ separated clusters  grows with the space dimension as $P(k)\sim \exp(-\alpha k^{ d/(d-1)})$ \cite{Aizenman97}. In our study, we take into account only one cluster per each disordered lattice constructed, in order to avoid presumable correlations of the data.

\begin{figure}[b!]
\begin{center}
\includegraphics[width=4cm]{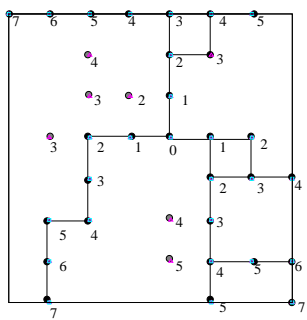}
\end{center}
\caption{\label{himia}
For all sites of a percolation cluster the chemical distance from the starting site is calculated. The minimal paths from all the sites on the edge of the percolation cluster to this starting point are found, which form the elastic backbone of the percolation cluster.}
\end{figure}

\begin{figure}
\begin{center}
\includegraphics[width=4cm]{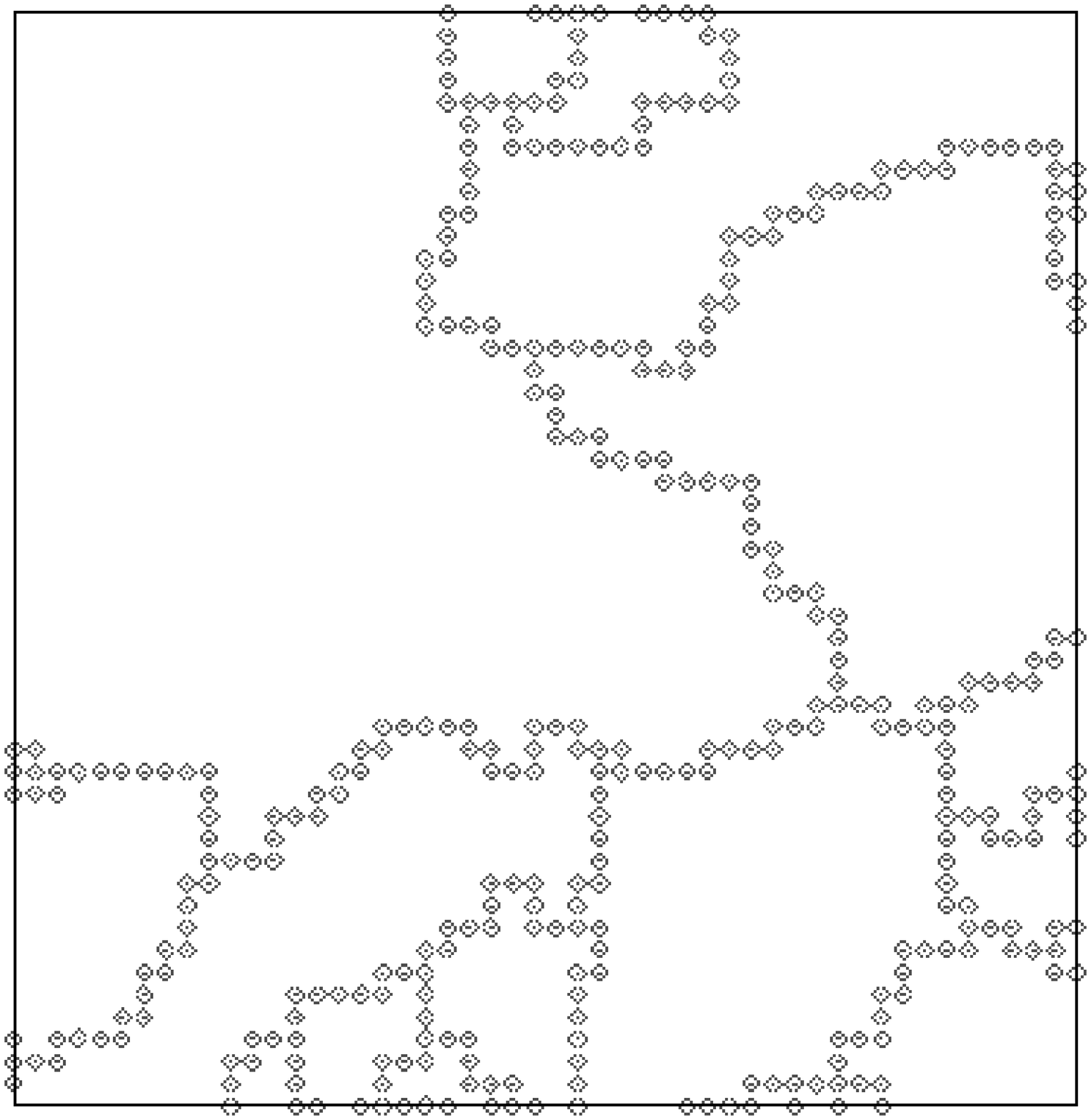}
\hspace*{2cm}
\includegraphics[width=4cm]{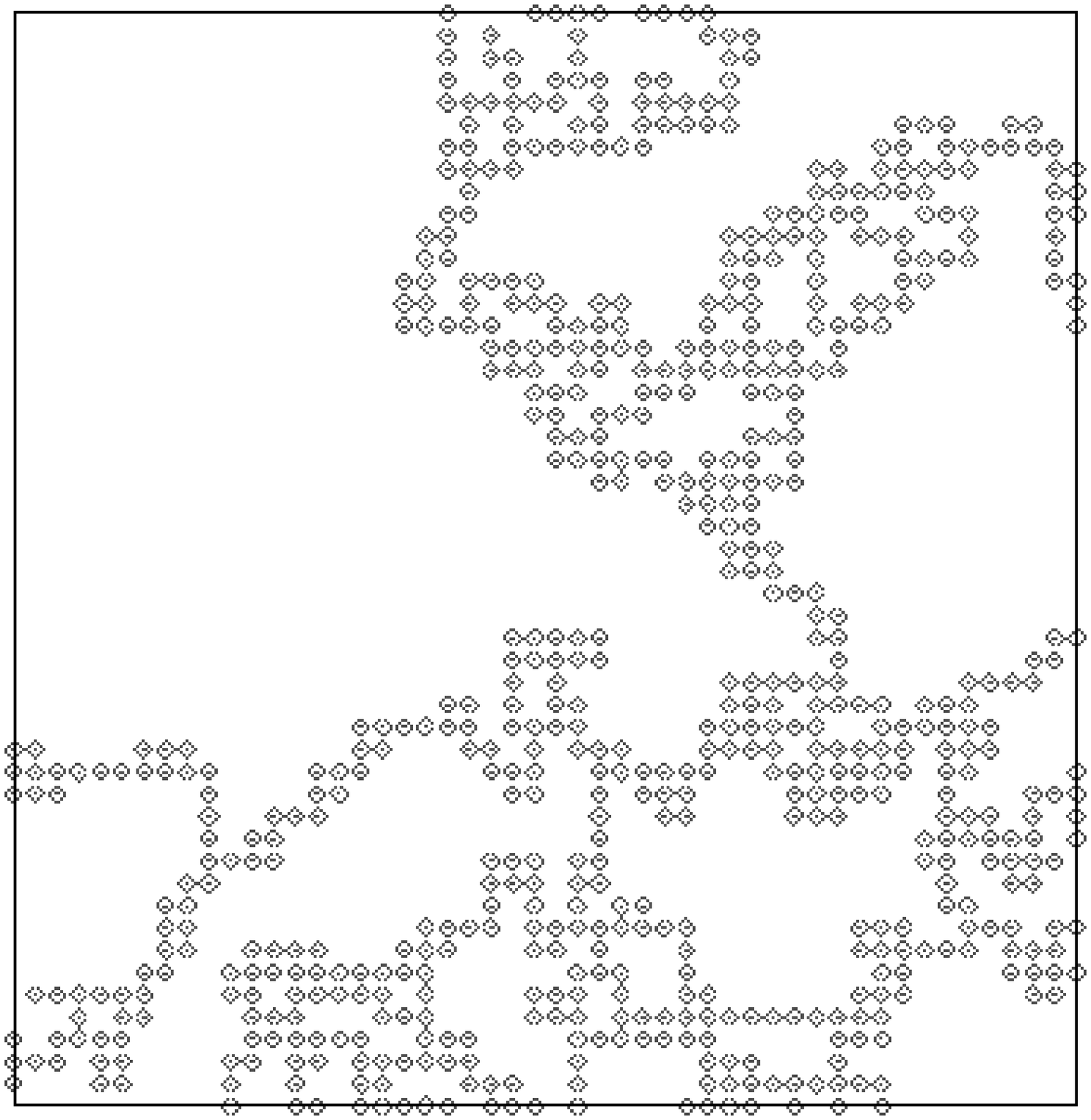}
\end{center}
\caption{\label{geom}Elastic and geometrical backbones of the percolation cluster depicted in Fig.~\ref{perc}.}
\end{figure}

Aiming to investigate the scaling of SAWs on percolative lattices, we are interested in the backbone of percolation clusters, 
which can be defined as follows. Assume that each bond (or site) of the cluster is
a resistor and that an external potential drop is applied at two ends of the cluster. The backbone is 
the subset of the cluster consisting of all bonds (or sites) through which the current flows; i.e., it is the structure left when 
all ``dangling ends" are eliminated from the cluster. The SAWs can be trapped in ``dangling ends",  therefore infinitely long chains can 
only exist on the backbone of the cluster. 
The algorithm for extracting  the  backbone of obtained percolation clusters was first introduced in Ref. \cite{Herrmann84} and improved in Ref. \cite{Porto97}.  
 We choose the starting point -- ``seed'' -- at the center of the cluster, and find the 
chemical distance $l$ of all the sites belonging to the percolation cluster to this starting point. In Fig.~\ref{himia}, the  starting point is numbered with 
 $0$, and  the chemical distance of all the other sites are depicted. The burning algorithm is divided into two parts. First, we start from some  site of the edge of the lattice  belonging to the percolation cluster and consider it as burning site. At the next step, if the nearest neighbor of this site has the chemical distance smaller than the burning site itself, the nearest neighbor site is burnt. Such a procedure ends when the ``seed"  site is reached. All the thus obtained burnt sites are located along 
the shortest path between the ``seed" and the given site at the edge of the percolation cluster, as is shown in Fig.~\ref{himia}. The same algorithm is 
applied successively to all the edge sites. 
As a result, we obtain the so-called skeleton or elastic backbone \cite{Havlin84}, shown in Fig.~\ref{geom}, left.
In the second part of the algorithm, we  consider successively each site of the elastic backbone, and check,
 whether a ``loop" starts from this site. ``Loop" is a path of sites, belonging to the percolation cluster, which is connected with the elastic backbone 
by no less than two sites. All sites of  the elastic backbone together with the sites of ``loops" form finally  the geometric backbone of the cluster (see Fig. \ref{geom}, right). 

 The results for fractal dimensions of the percolation cluster and its 
geometrical backbone in $d{=}2,3,4$ are compiled  in Table~\ref{dim}.

\section{Self-avoiding walks on percolation clusters}
\subsection{The method}

For the sampling of SAWs, we use the pruned-enriched Rosenbluth method (PERM), proposed in the work of Grassberger \cite{Grassberger97}.  The algorithm is based on ideas from the very first days of Monte Carlo simulations, the Rosenbluth-Rosenbluth (RR) method \cite{Rosenbluth55} and enrichment \cite{Wall59}. Let us consider the growing polymer chain, i.e. the $n$th monomer is placed at a randomly chosen neighbor site of the last placed $(n-1)$th  monomer ($n\leq N$, where $N$ is total length of polymer). The growth is stopped, if the total length $N$ of the chain is reached, then the next chain is started to grow from the same starting point. In order to obtain correct statistics, if this 
new site is occupied, any attempt to place a monomer at it results in discarding the entire chain. This leads to an exponentional ``attrition", the number of discarded chains grows exponentially with the chain length, which makes the method useless for long chains. 
In the RR method, occupied neighbors are avoided without discarding the chain, but the bias is corrected by means of giving a weight $W_n\sim (\prod_{l=1}^n m_l)$ to each sample configuration at the $n$th step 
, where $m_l$ is the number of free lattice sites to place the $l$th monomer.
When the chain of total length $N$  is constructed, the new one starts from the same starting point, until the desired number of chain configurations are obtained.  
The configurational averaging, e.g.,  for the end-to-end distance $r(N)\equiv \sqrt{R^2(N)}$,  then has the form:
\begin{eqnarray}
&&\langle r (N) \rangle=\frac{1}{Z_N}{\sum_{{\rm conf}}W_N^{{\rm conf}}(\vec{r}_N-\vec{r}_0)^2}, 
\,\,\,\,Z_N=\sum_{{\rm conf}} W_N^{{\rm conf}} \label {R},
\end{eqnarray}
where $\vec{r}_0$ is the position of the starting point of the growing chains, $\vec{r}_k$ is the position of the $k$th monomer,
and $Z_N$ is the partition sum.

The Rosenbluth method, however, also suffers from attrition: if all next neighbors at some step ($n<N$) are occupied, i.e., the chain is running into a ``dead end", the complete chain has to be discarded and the growth process has to be restarted. Combining the chain growth algorithm with population control, such as PERM (pruned-enriched Rosenbluth method) \cite{Grassberger97} leads to a considerable improvement of the efficiency by increasing the number of successfully generated chains. The weight fluctuations of the growing chain are suppressed in PERM by pruning configurations with too small weights, and by enriching the sample with copies of high-weight configurations \cite{Grassberger97}. These copies are made while the chain is growing, and continue to grow independently of each other. Pruning and enrichment are performed by choosing thresholds $W_n^{<}$
and $W_n^{>}$ depending on the estimate of the partition sum for the $n$-monomer chain. These thresholds are continuously updated as the simulation progresses. The zeroes iteration is a pure chain-growth algorithm without reweighting. After the first chain of full length has been obtained, we switch to $W_n^{<}$,
 $W_n^{>}$. If the current weight $W_n$ of an $n$-monomer chain is less than $W_n^{<}$, a random number $r={0,1}$ is chosen; if $r=0$, the chain is discarded, otherwise it is kept and its weight is doubled. Thus, low-weight chains are pruned with probability $1/2$. If $W_n$ exceeds  
$W_n^{>}$, the configuration is doubled and the weight of each identical copy is taken as half the original weight.  For a value of the weight  lying between the thresholds, the chain is simply continued without enriching or pruning the sample. 
For updating the threshold values we apply similar rules as in \cite{Hsu03,Bachmann03}: $W_n^{>}=C(Z_n/Z_1)(c_n/c_1)^2$ and $W_n^{<}=0.2W_n^{>}$, where $c_n$ denotes the number of created chains having length $n$, and the parameter $C$ controls the pruning-enrichment statistics. 
After a certain number of chains of total length $N$ is produced, the given tour  is finished and a new one starts. We adjust the pruning-enrichment control parameter such that on average 10 chains of total length $N$ are generated per each tour \cite{Bachmann03}. Also, what is even more important for efficiency, in almost all iterations at least one such a chain was created.  The number of different trajectories of  SAWs with $N$ steps can be then estimated as averaged weight:
\begin{equation}
 C_N  =\frac{1}{t}\sum_{{\rm conf}}W_N^{{\rm conf}},
\label{number}
\end{equation}
where $t$ is the number of successful tours.
PERM has been applied to a wide class of problems, in particular  study of $\Theta$-transition in homopolymers \cite{Grassberger97}, trapping of random walkers on absorbing lattices \cite{Mehra02}, study of protein folding \cite{Frauenkron98,Bachmann03} etc.   

\subsection{Results}


To study scaling properties of SAWs on percolating lattices, we have to perform two types of averaging:  the first average is performed over all SAW configurations on a single percolation cluster according to  (\ref{R});
the second average is carried out over different realizations of disorder, i.e. over all percolation clusters constructed:
\begin{eqnarray}
&&\overline{\langle r \rangle}{=}\frac{1}{M}\sum_{i{=}1}^M \langle r \rangle_i,\label{av}\\
&&\overline{ C_N }{=}\frac{1}{M}\sum_{i{=}1}^M  C_{N,i},
\end{eqnarray}
where $M$ is the number of different clusters and the index $i$ means that a given quantity is calculated on the cluster $i$.

\begin{figure}[t!]
\begin{center}
\includegraphics[width=7cm]{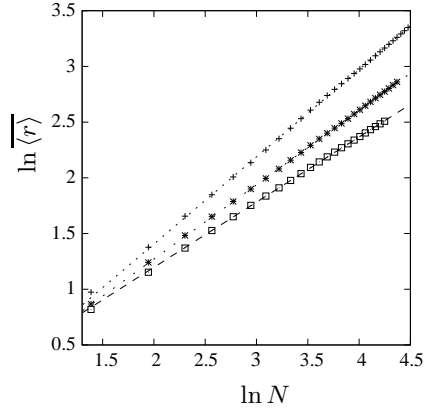}
\caption{Disorder averaged  end-to-end distance vs number of steps in double logarithmic scale for SAWs on the backbone of percolation 
clusters in 
 $d{=}2$ (pluses), $d{=}3$ (stars), $d{=}4$ (squares).
Lines represent linear fitting, statistical error bars are of the size of symbols.}
\label{sawr}
\end{center}
\end{figure}

\begin{table}[b!]
{\small
\caption{ \label{allnu} The exponent $\nu_{{\rm SAW}}^{p_c}$ for a SAW on a
percolation cluster. FL: Flory-like theories, EE: exact
enumerations, RS, RG: real-space and field-theoretic renormalization group, MC: Monte Carlo simulations. For SAWs on the regular lattice one has:
 $\nu_{{\rm SAW}}(d{{=}}2){=}3/4$~\cite{Nienhuis82}, $\nu_{{\rm SAW}}(d{{=}}3){=}0.5882(11)$~\cite{Guida98},
  $\nu_{{\rm SAW}}(d\geq 4){=}1/2$.}
\begin{indented}
\item[]
\begin{tabular}{@{}r l l  l  }
\br
 $\nu_{{\rm SAW}}^{p_c} \setminus d$  & 2 & 3 & 4 \\ 
  \mr  FL  Eq.~(\ref{kremer}) & 7/9 & 0.665& 0.594\\
 \cite{Sahimi84} & 0.778 & 0.662 & 0.593 \\
\cite{Roy90} & 0.77& 0.66 & 0.62 \\ 
\cite{Roy87}  & 0.770 & 0.656 & 0.57\\
 \cite{Aharony89} & 0.76 & 0.65 & 0.58\\
   EE 
\cite{Lam90} &0.745(10)& 0.635(10)&\\
 \cite{Rintoul94}&0.770(5)& 0.660(5)&\\
\cite{Ordemann00}&0.778(15)& 0.66(1)& \\
   RS
\cite{Sahimi84} & 0.778&0.724&\\ 
\cite{Lam84} & 0.767 & &  \\ 
 RG  
\cite{Blavatska04} & 0.785 & 0.678& 0.595 \\
\cite{Janssen07} & 0.796 & 0.669& 0.587 \\
 MC 
\cite{Woo91}
  & 0.77(1)  & &   \\
\cite{Grassberger93} & 0.783(3) & & 
\\
\cite{Lee96} & & 0.62--0.63 &0.56--0.57 
\\
\cite{Ordemann00}&0.787(10)& 0.662(6)&\\
our results  & $ 0.782\pm 0.003$ & $0.667\pm 0.003$ & $0.586\pm 0.003$ 
\\
 \br
\end{tabular}
\end{indented}
}
\end{table}

\begin{figure}[b!]
\begin{center}
\includegraphics[width=6.8cm]{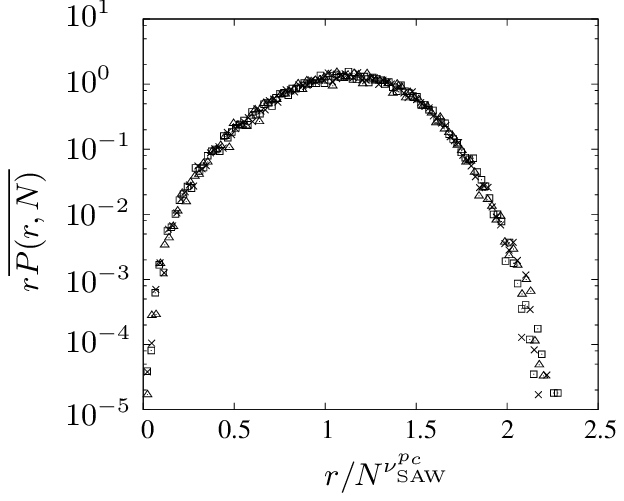}
\end{center}
\caption{Disorder averaged distribution function $r\overline{P(r,N)}$  vs the scaling variable $r/N^{\nu_{{\rm SAW}}^{p_c}}$ in $d{{=}}2$ dimensions. Lattice size $L{{=}}400$,
number of SAW steps $N{{=}}140$ (squares), $N{{=}}120$ (triangles), $N{{=}}100$ (crosses).}
\label{prsaw2}
\end{figure}

\begin{figure}[t!]
\begin{center}
\includegraphics[width=7cm]{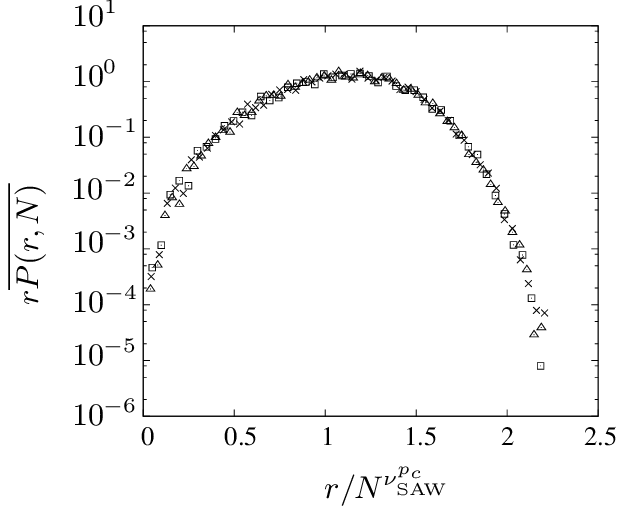}
\end{center}
\caption{Disorder averaged distribution function $r\overline{P(r,N)}$  vs the scaling variable $r/N^{\nu_{{\rm SAW}}^{p_c}}$ in $d{{=}}3$ dimensions. Lattice size $L{{=}}200$,
number of SAW steps $N{{=}}80$ (squares), $N{{=}}60$ (triangles), $N{{=}}50$ (crosses).}
\label{prsaw3}
\end{figure}

\begin{figure}[t!]
\begin{center}
\includegraphics[width=6.6cm]{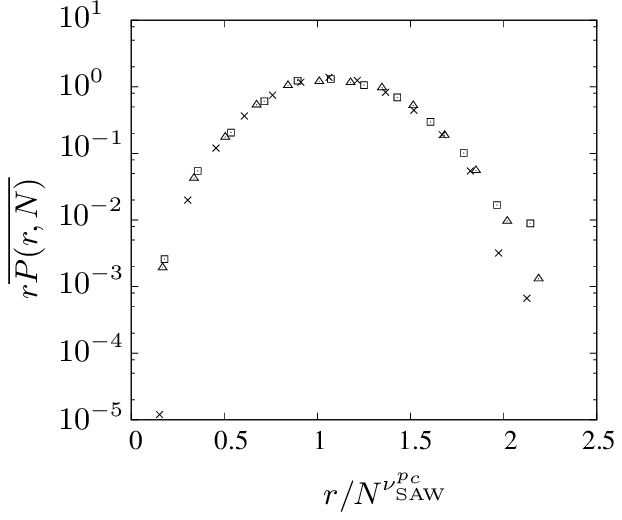}
\end{center}
\caption{Disorder averaged distribution function $r\overline{P(r,N)}$  vs the scaling variable $r/N^{\nu_{{\rm SAW}}^{p_c}}$ in $d{{=}}4$ dimensions. Lattice size $L{{=}}50$,
number of SAW steps $N{{=}}30$ (squares), $N{{=}}20$ (triangles), $N{{=}}15$ (crosses).}
\label{prsaw4}
\end{figure}

The SAW statistics crucially depends on the type of disorder averaging, namely, whether the disorder is considered to be ``annealed" (positions of defects are in thermodynamical equilibrium with the system) or ``quenched" (positions of defects are out of equilibrium). 
As it was pointed out in Ref.~\cite{Doussal91}, the correctness of results, obtained in the picture of ``quenched" disorder, depends on whether the location of the starting point of a SAW  is fixed 
while the configurational averaging is performed, or not. In the latter case, one has to average over all locations and effectively this corresponds to the case of annealed disorder.

An interesting question arises: what is the difference in statistics between SAWs walking on percolation clusters and the backbone of percolation clusters, after eliminating all the ``dead ends"? First Kim \cite{Kim90} claimed, based on a scaling argument, that the critical behavior on the percolation cluster is the same as
 that on the backbone, namely $\nu_{{\rm SAW}}^{p_c}=\nu_{{\rm SAW}}^{B}$. This 
equality was also assumed by Rammal \cite{Rammal84} in deriving a Flory formula for SAWs on fractal substrates. This can be easily explained: since the walks, which visit
 the dead ends are eventually terminated after a certain number of steps, the walks that survive in the limit $N\to \infty$ are those confined within the backbone. However, in a numerical study \cite{Woo91} it was found, that $\nu_{{\rm SAW}}^B>\nu_{{\rm SAW}}^{p_c}$, and, moreover, that $\nu_{{\rm SAW}}^{p_c}$ almost equals the value for SAWs on  pure lattices. 
It was argued, that the averaged end-to-end distance of SAWs on the backbone is significantly enhanced in comparison with averaging on the full percolation cluster.  We have checked this, comparing results obtained by us for the averaged end-to-end distance  $\overline{\langle r(N) \rangle}$ on percolation clusters and the backbone of percolation clusters. 
We conclude, that there is practically no difference in scaling for SAWs on both types of clusters,
the SAW statistics is  determined by the backbone of percolation clusters.

To study the scaling properties of SAWs on the backbone of percolation clusters, we choose as the starting point  the ``seed" of the cluster, and apply the  
PERM algorithm, taking into account, that a SAW can have its steps only on the sites belonging to the backbone of the percolation cluster. 
We use lattices of  size up to $L_{{\rm max}}{=}400, 200, 50$ in $d{=}2,3,4$, respectively, and performed averages over 
1000 percolation clusters in each case. Estimates for the critical exponents $\nu_{{\rm SAW}}^{p_c}$ were obtained by linear least-square 
fitting  (see Tables~\ref{2d}, \ref{3d}, \ref{4d} in the Appendix). Note, that since  we can construct lattices only of a finite size $L$, it is not possible to perform very long SAWs on it. 
For each $L$, the scaling for $\overline{\langle r(N) \rangle}$  holds  only up to some ``marginal" number of SAWs steps
$N_{{\rm marg}}\sim L^{1/\nu_{{\rm SAW}}^{p_c}}$ \cite {Blavatska08}.  We take this into account when analyzing the data obtained; 
for each lattice size we are interested only in values of $N<N_{{\rm marg}}$, thus avoiding distortions, caused by finite-size effects.
Our results for the scaling exponent $\nu_{{\rm SAW}}^{p_c}$ for SAWs on the backbone of percolation clusters \cite{Blavatska08} are given in Table~\ref{allnu} and
compared with previous estimates obtained by a variety of different techniques. 
 We see that the  value of $\nu_{{\rm SAW}}^{p_c}$ is larger than the  corresponding exponent on the pure lattice; 
presence of disorder leads to stretching of the trajectory of self-avoiding walks.

A simple modified Flory formula for the exponent of a SAW on a percolation cluster, proposed  a long time ago by Kremer \cite{Kremer81}, 
\begin{equation}
\nu_{\rm SAW}^{p_c}=3/(d_{p_c}^F+2),
\label{kremer}
\end{equation}
gives numbers in an astonishingly good agreement with our numerical data (see Table~\ref{allnu}). For the estimates we have used the values of 
the fractal dimension of percolation clusters from Table~\ref{dim}.  Since $d_{\rm up}=6$ for percolation and 
$d_{p_c}^F(d\geq6)=4$ \cite{Stauffer}, we receive from Eq.~(\ref{kremer}) that the exponent $\nu_{\rm SAW}^{p_c}(d\geq 6)=1/2$.
Note, that there exists a whole family of more sophisticated Flory-like theories
\cite{Kim90,Sahimi84,Rammal84,Kim87,Roy90,Roy87,Aharony89}.

The disorder averaged distribution function, defined via
\begin{equation}
\overline{\langle r \rangle}=\sum_{r}r {\overline {P(r,N)}}
\label{prob}
\end{equation}
 can be written in terms of the scaled variables 
$r/\overline{\langle r\rangle}$ as
\begin{equation}
r\overline{P(r,N)}\sim f(r/\overline{\langle r\rangle})\sim f(r/N^{\nu_{{\rm SAW}}^{p_c}}).
\end{equation}
The distribution function is normalized according to $\sum_{r}{\overline {P(r,N)}}{{=}}1$. The numerical results for the distribution function in $d=2$,$3$, and $4$ are 
shown in Figs. \ref{prsaw2}, \ref{prsaw3}, and \ref{prsaw4} for different $N$. When plotted against the 
scaling variable $r/N^{\nu_{{\rm SAW}}^{p_c}}$, the data are indeed found to nicely collapse onto a single curve, using our 
values for the  exponent $\nu_{{\rm SAW}}^{p_c}$ reported in Table~\ref{allnu}.

\begin{table}[t!]
\caption{ The connectivity constant $z^{p_c}$ for a SAW on a
percolation cluster. SS: series studies, EE: exact
enumerations, MC: Monte Carlo simulations. For SAWs on the regular lattice one has $z(d=2)=2.6385\pm0.0001$
\cite{Guttmann91}, $z(d=3)=4.68404\pm0.00009$ \cite{MacDonald00}, $z(d=4)=6.77507\pm0.00001$ \cite{MacDonald92}.}
\begin{indented}
\item[]
\begin{tabular}{@{}r lll }
\br $z^{p_c}\setminus d$  & 2 & 3 & 4 \\ 
\mr
 SS \cite{Barat91} & $1.31\pm0.03$ &&\\
  EE  
 \cite{Lam90}
&$1.53\pm 0.05$ & &   \\ 
\cite{Ordemann00} &
$ 1.565\pm0.005 $& $1.462\pm0.005$ &\\
  MC 
\cite{Woo91}
& $1.459\pm0.003$  &    &   \\
\cite{Ordemann00}
&
$1.456\pm0.005$ & $1.317 \pm 0.005$ &\\
 our results 
& $1.566\pm0.005$ & $1.459\pm0.005 $ & $1.340\pm0.005$ \\
$z\cdot p_c$ &1.564&1.460&1.333\\
\br
\end{tabular}
\label{allmu}
\end{indented}
\end{table}

\begin{table}[b!]
\caption{ \label{allgamma} The exponent $\gamma_{{\rm SAW}}^{p_c}$ for a SAW on a
percolation cluster. FL: Flory-like theories, EE: exact
enumerations, MC: Monte Carlo simulations. For a SAW on the regular lattice one has $\gamma_{\rm SAW}(d=2)=43/32$
\cite{Nienhuis84}, $\gamma_{\rm SAW}(d=3)=1.1596\pm 0.0002$ \cite{Guida98}, $\gamma_{\rm SAW}(d\geq4)=1$.}
\begin{small}
\begin{indented}
\item[]
\begin{tabular}{@{} r l l  l }
 \br
$\gamma_{{\rm SAW}}^{p_c} \setminus d$  & 2 & 3 & 4 \\ 
 \mr
 FL
\cite{Roy87} &
1.384 & 1.379& 1.27 \\
  EE 
 \cite{Lam90} &
$1.3\pm0.1 $  & &  \\ 
\cite{Ordemann00} &
$1.34\pm0.05$ & $1.29\pm0.05$  &\\
 MC 
\cite{Lee89} &
 $1.31\pm0.03$  & $1.40 \pm 0.02$   &   \\
 \cite{Ordemann00} &
$1.26\pm0.05$ & $1.19\pm 0.05$ &\\
 our results 
&  $ 1.350\pm 0.008$ & $1.269 \pm 0.008 $ & $1.250\pm0.008 $ \\
 \br
\end{tabular}
\end{indented}

\end{small}

\end{table}

\begin{figure}[t!]
\begin{center}
\includegraphics[width=8.2cm]{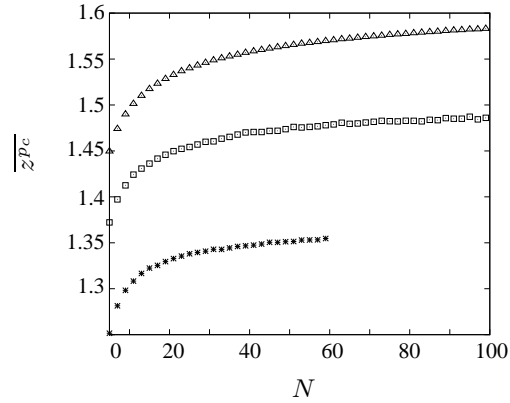}
\caption{Averaged connectivity constant for SAWs on the backbone of percolation 
clusters in  $d{=}2$ (triangles), $d{=}3$ (squares), $d{=}4$ (stars).}
\label{fugpc}
\end{center}
\end{figure}

\begin{figure}[b!]
\begin{center}
 \includegraphics[width=8cm]{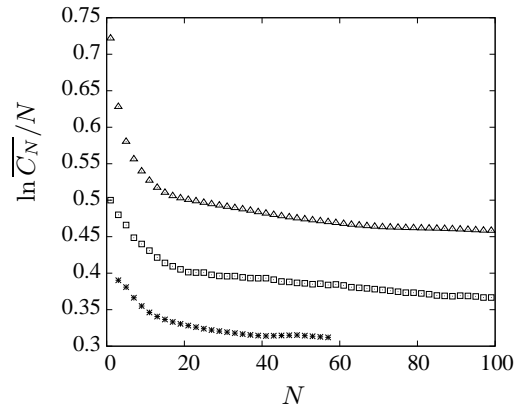}
\caption{Disorder averaged number of SAWs  configurations vs number of steps for SAWs on the backbone of percolation 
clusters in  $d{=}2$ (triangles), $d{=}3$ (squares), $d{=}4$ (stars).}
\label{gammapc}
\end{center}
\end{figure}
 
Let us now turn our attention to estimating the number of different possible SAW configurations ${\overline{ C_N }}$, defined by Eq.~(\ref{scaling}). First, let us note, that the fugacity or connectivity constant $z$ is obviously  affected by introducing disorder into the lattice. 
For the case, when the SAW is not confined only to the percolation cluster, namely when averaging over all the clusters is performed, then  $z^{p_c}{=}p_c z$ exactly. In Table~\ref{allmu} we present results of this estimate, taking values for $p_c$ from Table \ref{dim}. However, since each existing bond on the infinite percolation cluster has a non-trivial (correlated) probability of occurrence, a similar argument cannot be applied to the SAWs confined to the infinite percolation cluster only. However, 
enumeration estimates \cite{Chakrabarti83} suggested  $z^{p_c}{\simeq}p_c z$ up to $p_c$ for SAWs on the percolation cluster. It turns out, that this difference from linear dependence for incipient cluster is subtle and could hardly be detected. We have estimated $z_p$ as the averaged number of possibilities to perform the next step in the PERM 
procedure for SAWs on the backbone of percolation clusters (see Fig.~\ref{fugpc}); results are presented in Table~\ref{allmu}.

In the analytical study of Ref. \cite{Lyklema84}, it was argued that the exponent $\gamma$, governing the scaling 
of the number of SAW configurations, is not changed by the presence of disorder even at $p=p_c$. This 
was supported by an exact enumeration study \cite{Lam90}. However, this argument disagrees 
with results of studies \cite{Roy87,Ordemann00,Lee89}, where averaging over single percolation clusters was performed and different values for $\gamma_{{\rm SAW}}^{p_c}$ were found. 
In Ref. \cite {Roy87} it was proven, using scaling arguments, that at $p=p_c$ the exponents $\gamma_{{\rm SAW}}$  will crossover to
$\gamma_{{\rm SAW}}^{p_c}=\gamma_{{\rm SAW}}+d(\nu_{\rm SAW}^{p_c}-\nu_{{\rm SAW}})$ at $p=p_c$; the estimates based on this equality are given in the first row of 
Table~\ref{allgamma}. 

We obtained numerical estimates for $\gamma_{{\rm SAW}}^p$, studying the behavior of the quantity
\begin{equation}
\ln \overline{ C_N }/N=\ln(A)/N+\ln(z^{p_c})+(\gamma_{{\rm SAW}}^{p_c}-1)\frac{\ln N}{N},
\label{ggg}
\end{equation}
where $A$ is a constant. Figure \ref{gammapc} shows these values for the backbone of percolation clusters in $d=2,3,4$. Estimates for  $\gamma_{{\rm SAW}}^{p_c}$ are obtained by 
linear least-square fits (see Tables~\ref{g2d}, \ref{g3d}, \ref{g4d} in  the Appendix). Our final results are presented in Table~\ref{allgamma}.

\section{Random walks on percolation clusters}

\begin{figure}[b!]
\begin{center}
\includegraphics[width=7.8cm]{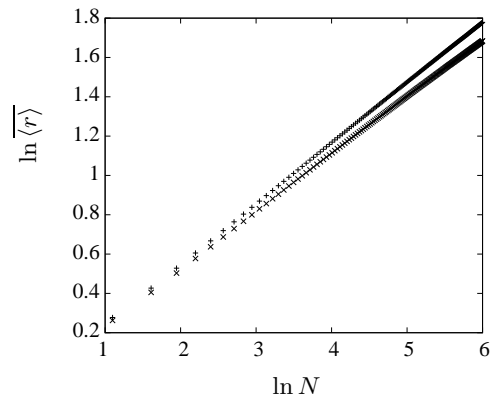}
\end{center}
\caption{\label{3drw} Averaged end-to-end distance vs number of steps in a double logarithmic scale of RWs on  percolation clusters (crosses) and the backbone of percolation 
clusters (pluses)  in $d=2$.}
\end{figure}

To simulate the diffusion process in a disordered medium, the picture of the ``ant in the labyrinth", proposed by de Gennes \cite{deGennes76} is traditionally used. Here the walker (an ``ant") starts at an arbitrary point on the diluted lattice and  tries to move randomly to the nearest site. If the randomly chosen direction leads to an empty site, it moves and the steps increment by 1. If the chosen site is occupied by a defect (in our case, does not belong to the percolation cluster) it stays at the current position for this time step. The growth is stopped, if the total number of steps  $N$ is performed, than the next trajectory is started to grow.

After averaging  the  end-to-end distance over RW configurations on a single percolation cluster, the  disorder average is carried out as in Eq.~(\ref{av}) over all constructed 
percolation clusters.
Let us note that, in contrast to the SAW problem, discussed previously, the scaling behavior of RWs on a percolation cluster is 
different from that on its backbone. Let us remind, that statistics of long SAWs on percolation clusters 
is nevertheless determined by its backbone, since the walks, which visit the ``dead ends" are eventually terminated after a certain number of steps. 
Simple random walks cannot be trapped in ``dead ends",
however, visiting these parts of a cluster will lead to some ``slowing down" of the diffusion process in comparison with 
the behavior on the  backbone where all the dead ends are removed. This is really confirmed by analyzing our results for the
averaged end-to-end distance of random walks on a percolation cluster and its backbone  (see Fig.~\ref{3drw}).  

\begin{figure}[b!]
\begin{center}
\includegraphics[width=6.4cm]{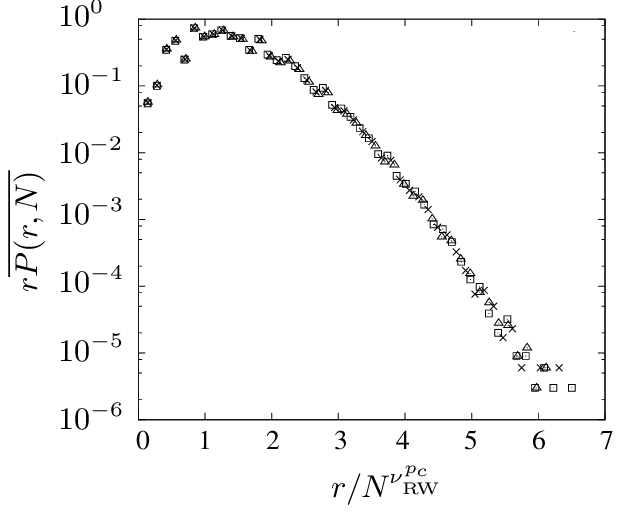}
\hspace*{0.5cm}
\includegraphics[width=6.8cm]{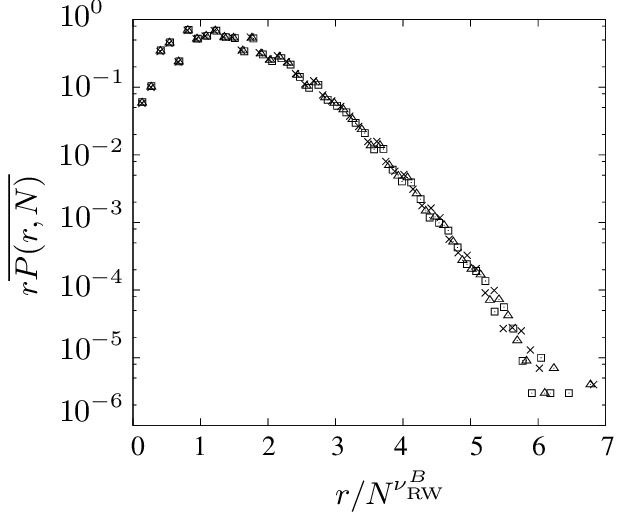}
\end{center}
\caption{Disorder averaged distribution function $r\overline{P(r,N)}$  vs the scaling variable $r/\overline{\langle r\rangle}$ in $d=2$ dimensions, left: percolation cluster, right: backbone of percolation cluster. Lattice size $L=400$,
number of RW steps $N{{=}}180$ (squares), $N{{=}}160$ (triangles), $N{{=}}140$ (crosses).}
\label{prrw2}
\end{figure}

\begin{figure}[t!]	
\includegraphics[width=7.1cm]{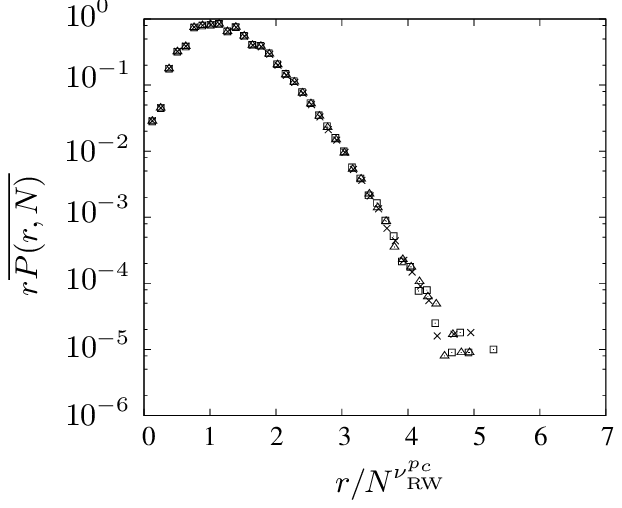}
\hspace*{0.5cm}
\includegraphics[width=6.58cm]{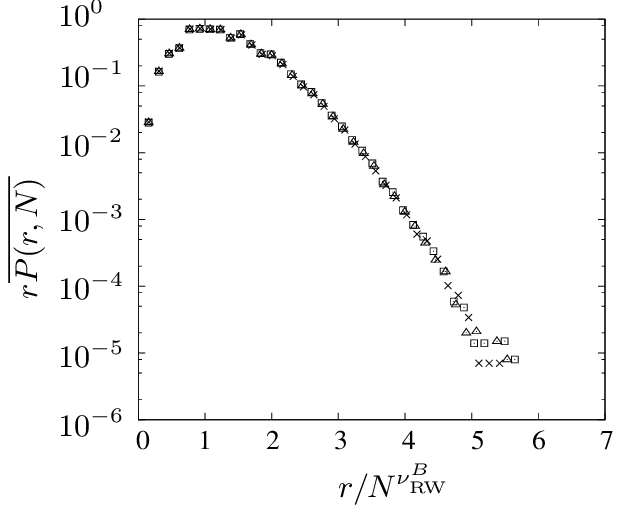}
\caption{Disorder averaged distribution function $r\overline{P(r,N)}$  vs the scaling variable $r/\overline{\langle r\rangle}$ in $d=3$ dimensions,
left: percolation cluster, right: backbone of percolation cluster.  Lattice size $L=200$,
number of RW steps $N{{=}}100$ (squares), $N{{=}}90$ (triangles), $N{{=}}80$ (crosses).}
\label{prrw3}
\end{figure}
\begin{figure}[t!]
\begin{center}
\includegraphics[width=6.8cm]{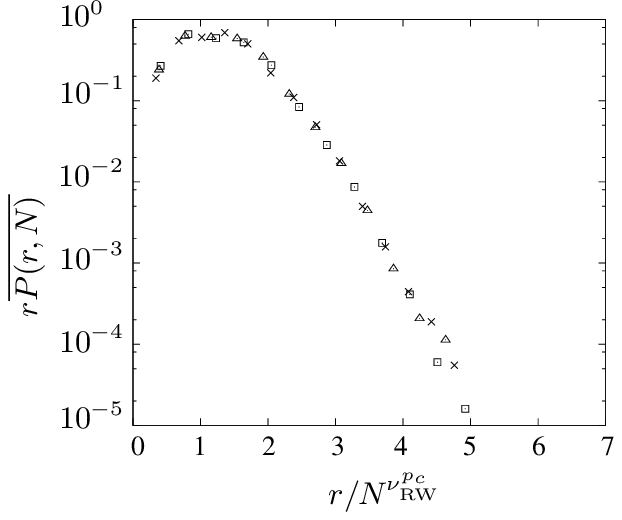}
\hspace*{0.45cm}
\includegraphics[width=7.1cm]{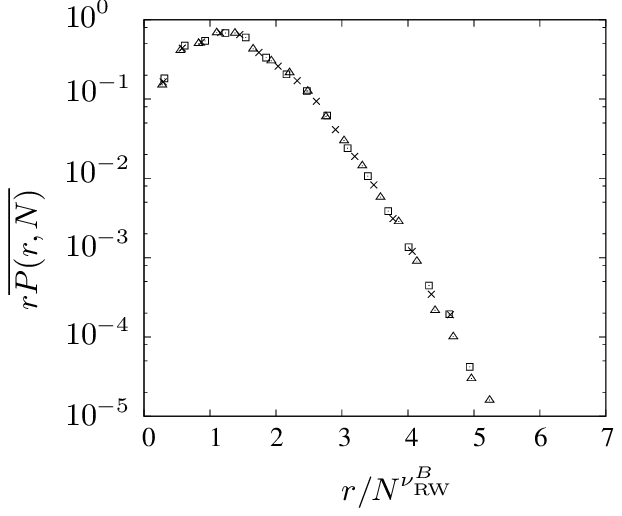}
\end{center}
\caption{Disorder averaged distribution function $r\overline{P(r,N)}$  vs the scaling variable $r/\overline{\langle r\rangle}$ in $d=4$ dimensions,
left: percolation cluster, right: backbone of percolation cluster.  Lattice size $L=50$,
number of RW steps $N{{=}}60$ (squares), $N{{=}}50$ (triangles), $N{{=}}40$ (crosses).}
\label{prrw4}
\end{figure}

We have studied RWs both on the percolation cluster and its backbone, 
performing  $10^7$ realizations on each cluster and average over $1000$ clusters in each space dimensions $d=2,3,4$. 
Estimates of scaling exponents $\nu_{{\rm RW}}^{p_c}$ and $\nu_{{\rm RW}}^{B}$, describing scaling of walks on percolation cluster and backbone,
 respectively, are obtained by linear least-square fitting and given  in Table~\ref{nurw}. One can see, that the inequality $\nu_{{\rm RW}}^{p_c} < \nu_{{\rm RW}}^{B}$ holds, and that the quantitative difference between these two values grows with increasing  the space dimension $d$. On the other hand, both values are {\em smaller} than the 
corresponding exponent $\nu_{{\rm RW}}=1/2$, governing scaling of random walks on the pure lattice: presence of disorder slows down the diffusion process.  The reason for this subdiffusive behavior is
 intuitively clear: due to the presence of defects,  the randomly walking particle  returns  back to already visited sites more often, thus its walking distance is    
 shorter than on the pure lattice. This has also been observed in recent studies of less disordered deterministic fractals such as two-dimensional Sierpinski carpet composites 
\cite{Anh07}.

The disorder averaged distribution function, defined in Eq.~(\ref{prob}),
and rewritten in terms of the scaled variables 
$r/\overline{\langle r\rangle}$ as:
\begin{equation}
r\overline{P(r,N)}\sim f(r/\overline{\langle r\rangle})\sim f(r/N^{\nu^{p_c}})
\end{equation}
is shown in Figs. \ref{prrw2}, \ref{prrw3} and \ref{prrw4} for $d=2,3,4$, both for the cases of percolation cluster and backbone. When plotted against the 
scaling variable $r/N^{\nu^{p_c}}$, the data are indeed found to nicely collapse onto a single curve, using our 
values for the  exponent $\nu_{{\rm RW}}^{p_c}, \nu_{{\rm RW}}^{B}$ reported in Table~\ref{nurw}.

\begin{table}[b!]
\begin{indented}
\item[]
\caption{\label{nurw} Exponents governing the scaling law of the end-to-end distance for RWs on percolation clusters and the backbone of percolation clusters. RS: real-space renormalization-group, EE: exact enumerations,  MC: Monte Carlo simulations.}
\begin{tabular}
{@{} r l l  l  l }
\br
$\nu_{{\rm RW}}^{p_c}\setminus d$  & 2 & 3 & 4  \\
\mr
 RS
\cite{Sahimi83}& 0.356 & 0.285 &\\
 EE 
\cite{Majid84} & 0.349$\pm0.002$ &&\\
\cite{Pandey83} & & 0.266$\pm0.01$ &\\

analytic \cite{Alexander82} & 0.352 & 0.268  &\\
 MC 
\cite{Avraham82} & $0.352\pm0.006$ & $0.271\pm0.004$ &\\
\cite{Havlin83} &$0.352\pm0.006$ &$0.271\pm0.004$&\\
\cite{Argyrakis84} & $0.392\pm 0.007$ & $0.282\pm0.003$ &\\
\cite{McCarthy88} & $0.348\pm0.009$ & $0.274\pm0.007$ &  \\
\cite{Lee00}& & & $0.222\pm0.007$  \\ 
our results &$ 0.353\pm 0.003 $&$ 0.273\pm 0.003$&$ 0.231 \pm 0.003$\\
\mr
$\nu_{{\rm RW}}^{B}\setminus d$  &2&3&4\\
\mr
analytic
\cite{Bug86} & $0.371\pm0.001$ &&\\
 EE
\cite{Hong84} & $0.372\pm0.005$ &&\\ 
MC \cite{Mastorakos93} &$0.370\pm0.003$ &&\\
our results &$0.372\pm 0.002 $&$ 0.306\pm0.002$ & $ 0.262\pm0.002$\\
\br
\end{tabular}
\end{indented}
\end{table}

\section {Conclusions}

We studied the scaling behavior of simple random walks and self-avoiding walks on disordered lattices. Both models are of great interest: RWs provide a good description of diffusion processes, SAWs  are successful in describing the universal properties of long flexible polymer macromolecules in a good solvent.  

We consider the case, when concentration $p$ of lattice sites allowed for walking equals the critical concentration $p_{c}$ and the incipient percolation cluster can be found in the system. 
Studying properties of percolative lattices, one encounters two possible statistical averages: 
in the  first, one considers  only percolation clusters with linear size much larger than the typical length of the physical phenomenon under discussion,  
the other statistical ensemble includes all the clusters, which can be found on a percolative lattice. 
In our study, we considered only the first case, being interested in random and self-avoiding walks on a percolation cluster, which has a fractal structure. 
In this regime, the critical behavior of both models belongs to a new universality class: scaling law exponents change  with the
dimension  of the (fractal) lattice on which the walk
resides. 

We performed numerical simulations of random and self-avoiding walks on percolation clusters and the backbone of percolation clusters on lattice sizes $L=400, 200, 50$ in space dimensions $d=2,3,4$,  respectively. Our results bring about the estimates for critical exponents, governing the scaling behavior of the models. We found that the statistics of SAWs is governed by the same scaling exponent both on a percolation cluster and its backbone: since the walks, which visit the dead ends are eventually terminated after a certain number of steps, the walks that survive in the limit $N\to \infty$ on a percolation cluster are those confined within its backbone. For 
simple random walks, however, the picture is different: they cannot be trapped in ``dead ends".
However, visiting these parts of a cluster will lead to some ``slowing down" of the diffusion process in comparison with the behavior on the backbone where 
all the dead ends are removed. We found that the inequality $\nu_{{\rm RW}}^{p_c} < \nu_{{\rm RW}}^{B}$ holds, and the quantitative difference between these two values grows with increasing  space dimension $d$. 

 The presence of disorder leads to a stretching of the trajectory of self-avoiding walks: the value of $\nu_{{\rm SAW}}^{p_c}$ is 
larger than the corresponding exponent on the pure lattice. However, the exponent $\nu_{{\rm RW}}^{p_c}$, governing scaling of random walks on percolative lattices is smaller than that on a pure lattice: presence of disorder slows down the diffusion process.  This can be explained as follows: due to the presence of defects,  the randomly walking particle  returns  back to already visited sites more often, thus its walking distance is    
 shorter than on the pure lattice. 

\section{Acknowledgments} Work supported in part by the German Science Foundation (DFG) through the Research Group FOR877. V.B. is grateful for support through the ``Marie Curie International Incoming Fellowship" EU Programme  and
to the Institut f\"ur Theoretische Physik, Universit\"at Leipzig,
for hospitality.

\section{Appendix}

To estimate the critical exponents $\nu_{{\rm SAW}}^{p_c}$, $\gamma_{{\rm SAW}}^{p_c}$,  linear least-square fits with varying 
lower cutoff for the number of steps $N_{{\rm min}}$ are used in order to detect possible corrections to scaling. 
For estimating $\nu_{{\rm SAW}}^{p_c}$ we use linear fits for the averaged end-to-end distance $\ln({\overline{\langle r(N) \rangle}})=\ln(A)+\nu_{\rm SAW}^{p_c}\ln N$, and for $\gamma_{{\rm SAW}}^{p_c}$
we employ Eq.~(\ref{ggg}).
Since this is an important aspect for assessing the quality of our final exponent estimates discussed in the main text, we have compiled in this Appendix 
these more detailed results in Tables~\ref{2d}-\ref{g4d}.
The $\chi^2$ value (sum of squares of normalized deviation from the regression line) divided by the number of degrees of freedom, DF, given in the last rows,
 serves as a test of the goodness of fit.
\newpage
\begin{table}[h!]
\caption{ Results  of linear fitting of obtained results for $\overline{\langle r\rangle}$ 
for SAWs in $d{=}2$ dimensions on the backbone of percolation clusters, $L{=}400$. }
\begin{indented}
\item[]
\begin{tabular}{@{} rlll}
\br
$N_{\rm min}$ & $\nu_{{\rm SAW}}^{p_c}$ & $A$ & $\chi^2/DF$\\ 
\mr 
11 & 0.790         $\pm$ 0.005  &  0.829         $\pm$ 0.003  & 2.396 \\
16 & 0.786         $\pm$ 0.005 & 0.841         $\pm$ 0.005 &  1.910 \\
21 &  0.782         $\pm$ 0.004    & 0.847         $\pm$ 0.005  &  1.479 \\
26 & 0.783         $\pm$ 0.003 & 0.842         $\pm$ 0.006 &  1.262 \\
31 & 0.782         $\pm$ 0.003  & 0.840           $\pm$ 0.007 & 0.839 \\
\br
\end{tabular}
\label{2d}
\end{indented}
\end{table}

\begin{table}[h!]
\caption{Same as Table \ref{2d} for $d{=}3$,  $L{=}200$.}
\label{3d}
\begin{indented}
\item[]
\begin{tabular}{@{}rlll}
\br
$N_{{\rm min}}$ & $\nu_{{\rm SAW}}^{p_c}$ & $A$ & $\chi^2/DF$ \\
\mr
11 &  0.668         $\pm$ 0.003 & 0.935         $\pm$ 0.004 & 2.269\\
16 &  0.669         $\pm$ 0.003 & 0.930         $\pm$ 0.004 & 2.054\\
21 &  0.669         $\pm$ 0.003 & 0.924         $\pm$ 0.004 &  1.345\\
26 &   0.667         $\pm$ 0.002   & 0.930         $\pm$ 0.006 & 0.743\\
31 & 0.668         $\pm$ 0.002 &  0.934         $\pm$ 0.008 & 0.844\\
\br
\end{tabular}
\end{indented}
\end{table}

\begin{table}[h!]
\caption{ Same as Table \ref{2d} for $d{=}4$, $L{=}50$.}
\begin{indented}
\item[]
\begin{tabular}{@{}rlll}
\br
 $N_{\rm min}$ & $\nu_{{\rm SAW}}^{p_c}$ & $A$ & $\chi^2/DF$\\ 
\mr
8 & 0.588         $\pm$ 0.002 &  1.025          $\pm$ 0.004 & 2.615\\
10 & 0.587         $\pm$ 0.002 & 1.023          $\pm$ 0.006 &  1.777 \\
12 &  0.586         $\pm$ 0.003 & 1.021          $\pm$ 0.01  & 0.978\\
14 & 0.586  $\pm$ 0.003 & 1.031          $\pm$ 0.01  & 0.767\\
\br
\end{tabular}
\label{4d}
\end{indented}
\end{table}

\begin{table}[h!]
\caption{ Results  of linear fitting of obtained results for ${\overline{ C_N }}$ 
for SAWs in $d{=}2$ dimensions on the backbone of percolation clusters, $L{=}400$. }
\begin{indented}
\item[]
\begin{tabular}{@{}rlll}
\br
 $N_{\rm min}$ & $\gamma_{{\rm SAW}}^{p_c}$ & $A$ & $\chi^2/DF$\\ 
\mr
16 &   $1.341   \pm                 0.005$ & $1.219     \pm             0.003$  & 3.135\\
21 &   $ 1.349   \pm               0.005$ & $1.189       \pm           0.003$ &  2.682 \\
26 &   $1.351       \pm           0.007$  & $1.168           \pm        0.002$ &   1.913\\
31 &   $1.352          \pm        0.008$ & $ 1.172    \pm              0.002$ & 1.621\\
36 &  $1.350               \pm    0.008 $ & $1.163         \pm         0.001$  & 0.704\\
\br
\end{tabular}
\label{g2d}
\end{indented}
\end{table}

\begin{table}[h!]
\caption{ Same as Table \ref{g2d} for $d{=}3$, $L{=}200$. }
\begin{indented}
\item[]
\begin{tabular}{@{}rlll}
\br
$N_{\rm min}$ & $\gamma_{{\rm SAW}}^{p_c}$ & $A$ & $\chi^2/DF$\\ 
\mr
11  &  1.265     $\pm$             0.004 &  1.82     $\pm$             0.003&  2.767 \\
16 &   1.268          $\pm$        0.005 & 1.192          $\pm$        0.003 & 2.135\\
21 &    1.270              $\pm$    0.006 & 1.184              $\pm$    0.003 &  1.968 \\
26 &   1.267                  $\pm$0.008  & 1.176                   $\pm$0.002 &   1.513\\
31 &   1.269                  $\pm$0.008   & 1.172          $\pm$        0.002 $\pm$ & 0.762\\
\br
\end{tabular}
\end{indented}
\label{g3d}
\end{table}

\begin{table}[h!]
\caption{Same as Table \ref{g2d} for $d{=}4$, $L{=}50$.}
\begin{indented}
\item[]
\begin{tabular}{@{}rlll}
\br
 $N_{\rm min}$ & $\gamma_{{\rm SAW}}^{p_c}$ & $A$ & $\chi^2/DF$\\ 
\mr
8  &  1.251     $\pm$             0.005 &  1.77     $\pm$             0.003&  1.767 \\
10 &   1.252          $\pm$        0.007 & 1.182          $\pm$        0.003 & 1.135\\
12 &    1.250              $\pm$    0.008 & 1.184              $\pm$    0.003 &  0.968 \\
\br
\end{tabular}
\end{indented}
\label{g4d}
\end{table}

\end{document}